\documentclass{Interspeech}

\interspeechcameraready

\title{Learning Annotation Consensus for Continuous Emotion Recognition}

\author[affiliation={1}]{Ibrahim}{Shoer}
\author[affiliation={1}]{Engin}{Erzin}

\affiliation{Department of Electrical and Electronics Engineering}{ Koc University}{Turkey}
\email{ishoer20@ku.edu.tr.edu, second@companyA.com, third@companyB.ai}

\usepackage{comment}
\usepackage[table]{xcolor}
\begin{document}

\maketitle

\begin{abstract}
    
In affective computing, datasets often contain multiple annotations from different annotators, which may lack full agreement. Typically, these annotations are merged into a single “gold standard” label, potentially losing valuable inter-rater variability. We propose a multi-annotator training approach for continuous emotion recognition (CER) that seeks a consensus across all annotators rather than relying on a single reference label. Our method employs a consensus network to aggregate annotations into a unified representation, guiding the main arousal-valence predictor to better reflect collective inputs. Tested on the RECOLA and COGNIMUSE datasets, our approach outperforms traditional methods that unify annotations into a single label. This underscores the benefits of fully leveraging multi-annotator data in emotion recognition and highlights its applicability across various fields where annotations are abundant yet inconsistent.

\end{abstract}

\section{Introduction}
Emotion Recognition (ER) technology aims to identify human emotions from various inputs like facial expressions, voice, and physiological signals. Effective ER is critical in fields such as human-robot interaction (HRI), healthcare, and virtual agents, where understanding and responding to human emotions are essential. Speech Emotion Recognition (SER) systems are particularly notable, classifying emotions based on both what is said (linguistic) and how it is said (paralinguistic). This duality captures a comprehensive emotional profile, crucial for applications that require nuanced emotional understanding \cite{schuller2018speech, ekman1987universals, russell1977evidence}.

The linguistic component benefits from advancements in automatic speech recognition and natural language processing, which have improved valence recognition \cite{sahu2019multi}. However, paralinguistic features are typically more reliable for detecting arousal and dominance and offer potential for language-independent ER \cite{calvo2010affect}. Despite considerable progress, challenges remain in SER, such as enhancing valence accuracy, ensuring robustness across varied real-world conditions, and addressing fairness in emotion recognition technologies \cite{triantafyllopoulos2023multistage, oates2019robust}

Understanding the affective content of speech is crucial for advancing artificial intelligence (AI) systems that can interpret and respond to human emotions. This capability is essential for developing robots and virtual agents capable of engaging in high-level human communication, enabling applications ranging from emotionally aware dialogue systems to assistive robotics in healthcare and education. For example, the phrase \textit{"Are you coming?"} can convey curiosity, impatience, or anger depending on the speaker's tone and delivery. Accurately interpreting such variations in speech is vital for AI-driven systems aiming to achieve natural human-like interactions \cite{gunes2017affective}.

In recent years, foundation models, including large-scale multimodal architectures, have demonstrated remarkable progress in natural language processing (NLP) and computer vision. However, these models still struggle with affect recognition due to the inherent subjectivity of human emotions and the lack of explicit emotional annotations in large-scale training data. Integrating emotion recognition into foundation models would enhance their ability to generate context-aware responses, improving their effectiveness in HRI \cite{poria2017review}. Continuous emotion recognition (CER) is also essential for robotics, where understanding emotional states can guide socially appropriate behavior, assist in therapeutic interventions, and improve human-AI collaboration in shared workspaces.

Beyond human-computer interaction, affective dimensions—such as valence and arousal—are instrumental in solving various interdisciplinary problems. One significant application is {video summarization}, where affective cues help identify emotionally salient moments, enabling the automatic creation of concise summaries that retain key emotional narratives \cite{9954146}. Emotion-based video summarization is particularly useful for content recommendation, film analysis, and accessibility tools for users with cognitive impairments. Additionally, in marketing and audience research, understanding viewers' emotional responses to video content can optimize advertisement placement and engagement strategies.

Despite the importance of affect recognition, current models often rely on a single, derived "gold standard" label as the ground truth, which oversimplifies the complexity of subjective emotions. Human annotators frequently disagree on emotional interpretations due to cultural backgrounds, personal experiences, and perceptual biases. Instead of collapsing annotations into a single reference label, we propose a {multi-annotator learning approach} that explicitly leverages all annotator inputs to infer a consensus representation. By incorporating a dedicated consensus module, our approach allows the model to learn from the full range of human perspectives, leading to improved emotion recognition performance.

In this work, we evaluate our proposed framework on two widely used emotion datasets: {RECOLA} \cite{ringeval2013introducing} and {COGNIMUSE} \cite{giannakopoulos2017cognimuse}. Our results show that training with multi-annotator annotations and a learned consensus yields higher performance compared to conventional single-label training methods. We demonstrate that preserving annotator diversity, rather than enforcing agreement, enables AI systems to develop a more robust and generalized understanding of affective states.

The remainder of this paper is structured as follows: Section~\ref{liter} reviews related research in affect recognition and consensus modeling. Section~\ref{sec:approach} describes our proposed framework for multi-annotator learning. Section~\ref{sec:expr} presents experimental results. Finally, Section~\ref{sec:conclusion} discusses the broader implications of our findings and directions for future research.

\section{Literature Review}
\label{liter}

Affective computing aims to enable machines to recognize, interpret, and respond to human emotions \cite{picard1997affective}. In this domain, the RECOLA and COGNIMUSE datasets have been widely used for evaluating automatic emotion recognition models. These datasets provide multimodal recordings annotated with affective labels and include individual annotations from all annotators, making them valuable benchmarks for evaluating emotion recognition algorithms and well-suited for our multi-annotator learning framework.
\begin{figure*}[t]
  \centering
  \includegraphics[width=\textwidth]{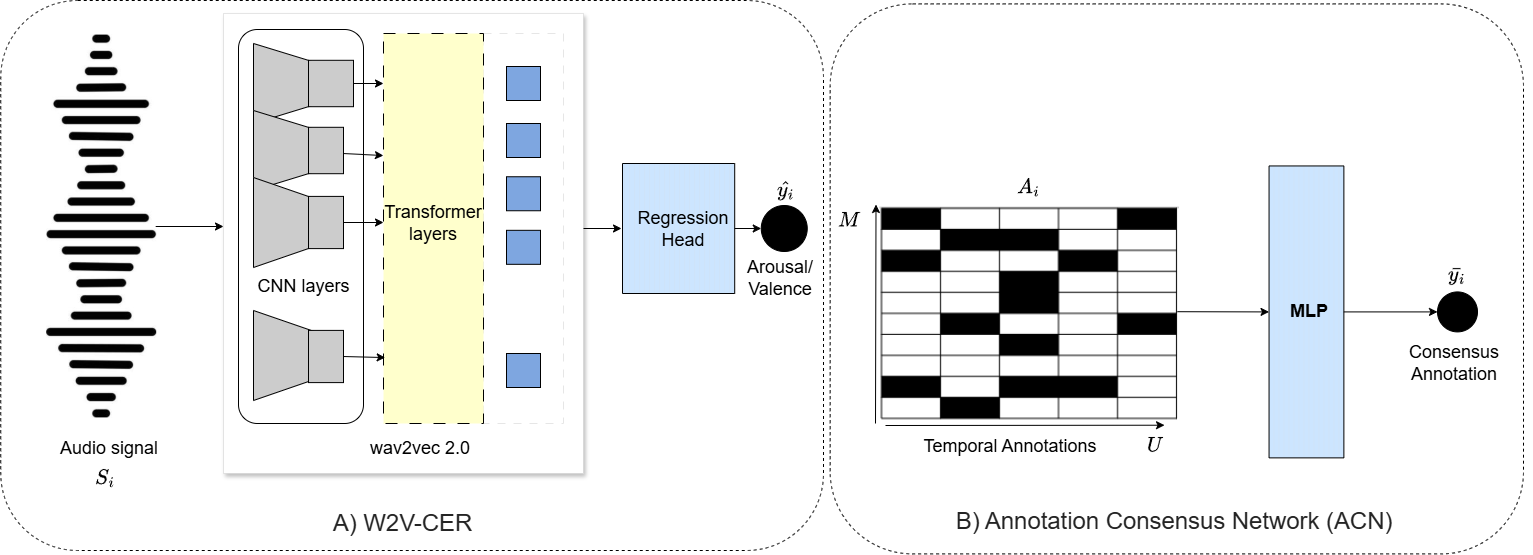}
  \caption{The architecture of the continuous emotion recognition network via learning annotation consensus consists of two integrated components: A) the baseline W2V-CER network, and B) the annotation consensus network (ACN), which incorporates multi-annotator learning.}
  \label{fig:speech_production}
\end{figure*}
The RECOLA (Remote Collaborative and Affective Interactions) dataset was introduced to study emotion recognition in a naturalistic setting where participants engaged in a collaborative task while being recorded via audio, video, and physiological signals \cite{ringeval2013introducing}. The dataset has been extensively used for continuous emotion recognition, particularly for valence and arousal modeling. Early approaches using RECOLA employed handcrafted features from speech and facial expressions \cite{ringeval2015prediction}. More recent studies leverage deep learning, including recurrent neural networks (RNNs) and transformers, to model the temporal dependencies of emotional expressions \cite{tzirakis2017end}. The dataset has also been used to evaluate multimodal fusion strategies, which combine complementary modalities to enhance affect recognition performance \cite{pan2020multimodalattentionspeechemotion}.

Similarly, the {COGNIMUSE} dataset was designed to study multimodal affect analysis in movie content \cite{giannakopoulos2017cognimuse}. It provides frame-level annotations of valence and arousal, along with other high-level semantic features such as saliency and event detection. The dataset has been used in various affective computing tasks, including emotion-based video summarization \cite{9954146}, where models identify emotionally significant moments to generate concise video representations. Recent works on COGNIMUSE have explored deep learning-based feature extraction methods.

Recent advancements have further enhanced emotion recognition capabilities. For instance, Wagner et al. (2023) explored transformer-based architectures in speech emotion recognition, achieving notable improvements in valence prediction without explicit linguistic information \cite{wagner2023dawn}. Additionally, Hayat et al. (2021) proposed a multitask learning approach to model emotions evoked by movies, emphasizing the importance of considering individual annotator responses alongside aggregated data \cite{hayat2021recognizing}.

Despite progress in affective computing with these datasets, a significant challenge remains: dealing with subjectivity in emotion annotations. People perceive emotions differently, leading to low agreement between annotators and raising concerns about the reliability of using a single “gold standard” label.

Emotion perception is inherently subjective, making it difficult to establish a definitive ground truth for affective computing tasks. Traditional approaches to emotion annotation involve aggregating multiple annotators’ ratings into a single consensus label, often through simple statistical methods such as mean or median computation. However, such methods ignore valuable information about individual differences in emotional perception \cite{calvo2015oxford}.

Several studies have explored methods to account for annotator disagreement in emotion recognition. One approach involves {weighted aggregation}, where annotators are assigned different weights based on their reliability and consistency \cite{grimm2007primitives}.

In the case of RECOLA and COGNIMUSE, multiple studies have highlighted the low agreement among human annotators. In RECOLA, annotators exhibited higher agreement for arousal compared to valence, likely due to the more observable nature of arousal-related features such as speech intensity and pitch variation \cite{ringeval2013introducing}. Similarly, COGNIMUSE annotations reveal variations in perceived valence and arousal across different movie genres and audiences \cite{giannakopoulos2017cognimuse}. These findings suggest that simple aggregation techniques may not be sufficient for training robust emotion recognition models.

By integrating multi-annotator learning with deep neural networks, we can design systems that learn a consensus representation while preserving valuable information about annotator disagreement. In this study, we propose a consensus-based training framework for affect recognition using RECOLA and COGNIMUSE. Our approach explicitly incorporates all annotator responses, allowing the model to learn from diverse perspectives and improve emotion recognition performance. Our results demonstrate that training with a consensus network outperforms traditional gold standard methods, highlighting the importance of preserving annotator variability in affective computing.

\section{Proposed Methodology}
\label{sec:approach}

Our primary objective is to predict two core dimensions of affect—arousal and valence—from raw audio signals using the wav2vec 2.0 architecture, which we fine-tune for affective computing tasks. Following recommendations in \cite{wagner2023dawn}, we freeze the convolutional feature encoder layers while fine-tuning the higher-level transformer blocks. This approach leverages the robust, pre-trained audio representations in the lower layers and focuses the training updates on the transformer layers that are more directly responsible for capturing fine emotional details.

The continuous emotion recognition network, denoted as \emph{W2V-CER}, processes a batch of $N$ audio signals $\{S_1, S_i, ..., S_N\}$ through the frozen CNN layers to obtain latent representations, which are then enriched contextually by the trainable transformer layers. A regression head maps each contextually enriched embedding to predicted outputs $\{\hat{y}_1, \hat{y}_i, ..., \hat{y}_N\}$, assessing the arousal and valence dimensions.

In addition to the main CER model, we introduce the \emph{Annotators Consensus Network (ACN)} to address the inherent subjectivity in emotional annotations from multiple annotators. For a given audio segment annotated by $U$ annotators, each providing $M$ temporal annotations, the resulting inputs are represented as $\{A_1, A_i, ..., A_N\}$ with a shape of $M \times U$ for each segment. The ACN processes these annotations through an MLP to produce consensus outputs $\{\bar{y}_1, \bar{y}_i, ..., \bar{y}_N\}$ on the emotional states, reducing variability and improving the generalizability of the predictions.

The architecture of both networks is depicted in Figure \ref{fig:speech_production}. The models are trained for either arousal or valence. We tried to jointly train arousal and valence, an improvement was only noticed on the Recola dataset for valence, we report the results of the joint training on Recola in the results section \ref{sec:expr}.
\subsection{Loss Function}
To train our models, we employed the Concordance Correlation Coefficient (CCC) loss, which is particularly well-suited for tasks where maintaining agreement between predicted and actual values is crucial. The CCC loss is formulated as follows:
\begin{equation}
    L_{CCC}(x, y) = 1-\frac{2 \rho \sigma_x \sigma_y}{\sigma_x^2 + \sigma_y^2 + (\mu_x - \mu_y)^2}
\end{equation}
where $\rho$ is the Pearson correlation coefficient between the predictions and the ground truth, providing a robust metric that combines measures of accuracy and precision.

\subsection{CER Models}
The baseline CER model, W2V-CER, is trained with the $L_{CCC}(y, \hat{y})$ loss function that aligns the W2V-CER predictions with the ground truth annotations.

Our proposed CER model, W2V-CER-ACN, is a combination of W2V-CER and ACN networks that are jointly trained with the weighted loss,  $L_{CER-ACN}$, given as
\begin{equation}
    L_{CER-ACN} = \alpha \cdot L_{CCC}(y, \bar{y}) + \beta \cdot L_{CCC}(\bar{y}, \hat{y}),
\end{equation}
where $\alpha$ and $\beta$ are hyperparameters that balance the influence of two loss functions. 
The loss function  $L_{CCC}(y, \bar{y})$  measures the discrepancy between the ground truth annotations and the learned consensus from the ACN. The second loss,  $L_{CCC}(\bar{y}, \hat{y})$, evaluates the alignment between the learned consensus and the predictions of the W2V-CER model. This dual-loss framework ensures that the model accurately captures true annotations while maintaining consistency with the consensus derived from multiple annotators’ perspectives.

By integrating both the W2V-CER and the ACN in this manner, our system is robust to the subjectivity of individual annotators and reduces model complexity by minimizing the number of trainable parameters. This configuration leads to stable convergence and lower computational overhead, which is particularly advantageous in large-scale applications.

\section{Experimental Evaluation}
\label{sec:expr}
\subsection{Datasets}
We utilized the RECOLA and COGNIMUSE datasets, which are noted for their comprehensive multimodal emotional annotations, including multiple annotators for arousal and valence. The RECOLA dataset consists of 9.5 hours of multimodal recordings from 46 French-speaking participants and includes annotations for arousal and valence made by three males and three females, providing a balanced perspective on affective states. This dataset is widely used in the research community, particularly highlighted in the AVEC 2016 challenge that featured 9 subjects for training and 9 for validation. For our validation, we adopted the gold standard annotations from AVEC 2018 to ensure comparability and consistency with previous studies. RECOLA also offers a balanced gender distribution and continuous annotations for valence and arousal at 25~Hz frequency.

The COGNIMUSE dataset features 30-minute segments from 7 Academy Award-winning films, annotated for both intended (used as the ground truth) and experienced emotions (used as input to the ACN), providing a unique dataset for evaluating models on both predicted and perceived emotional states. Each of the 7 participants attended two annotation sessions, enhancing the dataset with varied emotional insights. Annotations in COGNIMUSE are similarly detailed, provided at 25~Hz frequency, which facilitates a fine-grained analysis of emotional trajectories over time.

\subsection{Training}
Our training protocol was designed to evaluate the performance differences between the baseline W2V-CER and the proposed W2V-CER-ACN models. We conducted training over 15 epochs, employing a learning rate of $5 \times 10^{-4}$ and a batch size of 32. In terms of temporal resolution, data from RECOLA was segmented into 3-second windows with a 0.4-second shift, allowing us to capture transient emotional expressions effectively. For COGNIMUSE, we used 5-second windows with a 3-second shift to accommodate the longer-duration emotional turns typical of cinematic content.

To ensure rigorous and comprehensive testing, we implemented a 7-fold cross-validation approach on the COGNIMUSE dataset, testing the model on each film individually while training on the remaining films. During the joint training of our networks, we set both hyperparameters $\alpha$ and $\beta$ to 0.5 to equally balance the importance of aligning predictions with the ground truth annotations and achieving consensus among the annotators' experienced emotions.

\subsection{Results}

Our experimental evaluations on the RECOLA and COGNIMUSE datasets aimed to assess the impact of integrating the Annotators Consensus Network (ACN) on model performance, with a specific focus on the prediction accuracy of arousal and valence. We measured performance using the Concordance Correlation Coefficient (CCC), where higher values indicate better alignment with the ground truth annotations.
\begin{table}[h]
    \centering
    \caption{CCC performance of continuous emotion recognition on the RECOLA dataset for individually and jointly trained valence and arousal prediction tasks.}
    \begin{tabular}{lcc}
        \toprule
        Task & W2V-CER & W2V-CER-ACN \\
        \midrule
        Valence & 0.391       & \cellcolor{yellow!50}0.437 \\
        Arousal & 0.651       & \cellcolor{yellow!50}0.666 \\
    Valence \& Arousal    & 0.438/0.645 & \cellcolor{yellow!50}0.482/0.657 \\
    \bottomrule
    \end{tabular}

    \label{tab:recola_results}
\end{table}




The CCC score results on the RECOLA dataset are presented in Table~\ref{tab:recola_results}. The proposed W2V-CER-ACN model shows a significant improvement in valence prediction, achieving a CCC score of  $0.437$  compared to  $0.391$  for the baseline W2V-CER model. Although the improvement is more modest, the W2V-CER-ACN model shows better performance in arousal prediction, achieving a CCC score of  $0.666$  compared to  $0.651$  for the baseline W2V-CER model.

The last row in Table~\ref{tab:recola_results} shows the CCC score performances of CER models trained jointly for predicting valence and arousal attributes. The jointly trained CER models consistently show improved performance with the use of ACN, while joint training generally enhances valence prediction accuracy. For valence and arousal predictions with the jointly trained W2V-CER-ACN, the CCC scores are $0.482/0.657$ compared to $0.438/0.645$ for the baseline W2V-CER model. This improvement indicates a significant enhancement in the model's overall ability to interpret and predict emotional states accurately.

These results highlight the ACN's effectiveness in refining the model's predictions through the utilization of a consensus from multiple annotations, ensuring closer alignment with the true emotional states represented in the dataset.

The results underscore the ACN's role in enhancing emotional prediction performance in affective computing models. By incorporating annotations from multiple sources and learning a consensus, the ACN improves CCC scores and enhances model robustness across diverse and complex emotional datasets. This capability is particularly crucial for applications requiring dynamic content adaptation and interactive systems where more accurate emotion recognition is essential. Moreover, the consistent improvements across both datasets and in both emotional dimensions (arousal and valence) confirm the ACN's adaptability and effectiveness.

\section{Conclusion}
\label{sec:conclusion}
The integration of the ACN into affective computing models presents a promising avenue for future research, particularly exploring its applicability across other multimodal datasets and its potential in real-time emotional state prediction. These capabilities could significantly impact interactive technologies and empathy computing, facilitating more nuanced and responsive interactions. The challenges observed in valence prediction in the COGNIMUSE dataset highlight the need for further refinements in handling films with varied emotional content, suggesting directions for future enhancements to improve model accuracy and adaptability in diverse real-world scenarios.

Furthermore, the ACN presents a promising avenue for future research and application beyond emotional state prediction. The consensus mechanism utilized by the ACN could be adapted for other multimodal tasks such as conflict resolution in collaborative environments, crowd-sourced decision-making processes, or even in medical diagnostic systems where multiple expert opinions are beneficial. Each of these applications could benefit from a model that synthesizes diverse inputs into a coherent output, potentially enhancing decision-making accuracy and reliability.

\bibliographystyle{IEEEtran}
\bibliography{mybib}

\end{document}